\documentclass[1p]{elsarticle}
\usepackage{amssymb}

\newcommand{\clG}{{\cal G}}
\newcommand{\clP}{{\cal P}}
\newcommand{\clF}{{\cal F}}
\newcommand{\clC}{{\cal C}}
\newcommand{\mcT}{\mathcal{T}}
\newcommand{\clM}{\mathcal{ M}}

\newcommand{\clL}{{\cal L}}
\newcommand{\hh}{\tilde{h}}
\newcommand{\hL}{\hat{L}}
\newcommand{\hC}{\hat{C}}
\newcommand{\ha}{\hat{a}}
\newcommand{\hH}{\hat{\cal H}_0}
\newcommand{\hclL}{\hat{\cal L}}
\newcommand{\PBr}{\{H,\dots\}_{PB}}
\newcommand{\clE}{\mathcal{E}}

\newcommand{\prt}{\partial}

\newcommand{\rgl}{\rangle}
\newcommand{\lgl}{\langle}

\newcommand{\vep}{\varepsilon}
\newcommand{\bt}{\bar{\tau}}
\newcommand{\be}{\begin{equation}}
\newcommand{\ee}{\end{equation}}
\newcommand{\bea}{\begin{eqnarray}}
\newcommand{\eea}{\end{eqnarray}}

\begin{document}

\begin{frontmatter}

\title{Subdiffusion in classical and quantum
nonlinear Schr\"odinger equations with disorder}

\author{A. Iomin\corref{cor}}
\ead{iomin@physics.technion.ac.il}

\cortext[cor]{Corresponding author}
\address{Department of Physics, Technion, Haifa, 32000, Israel}

\begin{abstract}
The review is concerned with the nonlinear Schr\"odinger equation (NLSE) in
the presence of disorder. Disorder leads to localization
in the form of the localized Anderson modes (AM), while nonlinearity
is responsible for the interaction between the AMs and transport.
The dynamics of an initially localized wave packets
are concerned in both classical and quantum cases.
In both cases, there is a subdiffusive spreading, which is explained in
the framework of a continuous time random walk (CTRW), and it is shown
that subdiffusion is due to the transitions between those
AMs, which are strongly overlapped.
This overlapping being a common feature of both classical and
quantum dynamics, leads to the clustering with an effective 
trapping of the wave packet inside each cluster.
Therefore, the dynamics of the wave packet corresponds to the
CTRW, where the basic mechanism of subdiffusion is an entrapping
of the wave packet with delay, or waiting, times
distributed by the power law $w(t)\sim 1/t^{1+\alpha}$, where $\alpha$ is
the transport exponent. It is obtained that $\alpha=1/3$ for the classical NLSE
and $\alpha=1/2$ for the quantum NLSE.


\end{abstract}


\end{frontmatter}

\section{Introduction}\label{sec:int}
\def\theequation{\thesection.\arabic{equation}}
\setcounter{equation}{0}

It is well known that wave propagation in random media can be
described in the framework of the Fokker-Planck equation, under
certain conditions \cite{davis}. In modern optical experiments with
nonlinear media a suitable description can be developed in the
framework of the fractional kinetics based on fractional
integro-differentiation. This concept of differentiation of
non-integer orders arises from works of Leibniz, Liouville,
Riemann, Grunwald, and Letnikov, see \textit{e.g.},
\cite{oldham,podlubny,SKM}. Its application is related to random
processes with power law distributions. The latter corresponds to the
absence of characteristic average values for processes exhibiting
many scales \cite{klafter,shlesinger}.

A typical example of fractional dynamics in optics is a realization of
a competition between localization and nonlinearity that leads to
anomalous transport \cite{Sh93,molina,fks,PS,Iomin,mulansky,fks1}. This dynamics
is described in the framework of the nonlinear Schr\"odinger
equation (NLSE) in the presence of an external field $V=V(x),
~x\in(-\infty,+\infty)$. The wave spreading, described by the
wave function, is governed by the NLSE in the presence of disorder
\be\label{am1}
 i\prt_t\psi=-\prt_x^2\psi+\beta|\psi|^2\psi+V\psi\, ,
\ee %
where $\beta$ is a nonlinearity parameter. The variables are
chosen in dimensionless units and the Planck constant is
$\hbar=1$. The random potential $V=V(x), ~x\in(-\infty,+\infty)$
is such that for the linear case $(\beta=0)$ Anderson
localization takes place \cite{Anderson,LGP}, and the system is
described by the exponentially localized Anderson modes (AM)s
$\Psi_{\omega_k}\equiv\Psi_k(x)$, such that
\begin{equation}\label{eq_AM}
[-\prt_x^2+V(x)]\Psi_k(x)=\omega_k\Psi_k(x)\, ,
\end{equation}
where $\Psi_{\omega_k}(x)$ are real functions and the eigenspectrum
$\omega_k$ is discrete and
dense \cite{LGP}. The problem in question is an evolution of an
initially localized wave function $\psi(t=0)=\psi_0(x)$.
It can be also a stationary state of the NLSE, which is localized
with the same Lyapunov exponent as in the AMs \cite{IF2007,FIM}.

This problem is relevant to experiments in nonlinear optics, for
example disordered photonic lattices \cite{f1,lahini}, where
Anderson localization was found in the presence of nonlinear
effects, as well as to experiments on Bose-Einstein Condensates in
disordered optical lattices
\cite{BECE1,BECE3,SanchAspect,aspect}. A discrete analog of
Eq. (\ref{am1}) is extensively studied
numerically \cite{Sh93,molina,fks,PS,fks1}, and a subdiffusive
spreading of the initially localized wave packets has been
observed with the mean squared displacement (MSD)
$\lgl x^2(t)\rgl=\int|\psi(t)|^2x^2dx\sim
t^{\alpha}$, where a transport exponent $\alpha$ was found to be $2/5$
\cite{PS} and $1/3$ \cite{fks}. This difference has been explained in
\cite{MI2012}, where it has been shown that the former result is
for the Markovian subdiffusion due to the range-dependent diffusion
coefficient, while the latter one corresponds to non-Markovian
fractional diffusion of a percolation type.

Subdiffusion of wave packets was also obtained
analytically \cite{Iomin,iom1,MI2012} in the limit of the large times
asymptotic. In that case the
dynamics of the wave packet has been approximated by the
fractional Fokker-Planck equation (FFPE) due to the arguments of a
so-called continuous time random walk (CTRW).

The concept of the CTRW was originally developed for mean first
passage time in a random walk on a lattice with further
application to a semiconductor electronic motion \cite{montweiss}.
The mathematical apparatus of the fractional CTRW is well
established for many applications in physics, see {\it e.g.},
\cite{klafter,shlesinger,bouchaud,zaslavsky,benAvrHavlin}.

In the present work we concern with the
physical mechanism of this subdiffusion
obtained in Refs. \cite{Iomin,iom1,MI2015,Iomin2013} and give
a new insight of the subdiffusion transport exponent $\alpha$,
related to both classical and quantum
properties of the nonlinear interaction term in Eq. (\ref{am1}).
To this end we consider the NLSE (\ref{am1}) and its quantum counterpart
(quantum NLSE), when the wave functions $\psi(x)$ are considered as
operators $\hat{\psi}(x)$ satisfying to the commutation rule
$[\hat{\psi}(x),\hat{\psi}^{\dag}(x')]=\delta(x-x')$ \cite{QNLSE,bikt1986}.
The paper consists of two parts. The first one is devoted to
the classical analysis, which is based on
mapping the nonlinear Eq. (\ref{am1}) onto the linear Liouville
equation for the probability amplitude $|\psi(x,t)|^2$, where the
transition elements in the Liouville operator are determined by
the nonlinear term in Eq. (\ref{am1}). Therefore, we proceed by
developing the CTRW
approach for the corresponding Liouville equation \cite{kenkre}.
The second part of the review is devoted to the quantum NLSE, which a quantum
counterpart of Eq. (\ref{am1}). The quantum analysis is based on mapping
the quantum system on the basis of coherent states, and in the framework
of the obtained equations we study four-modes decay processes and develop
a quantum CTRW and construct a generalized master equation as a quantum 
counterpart of the Liouville equation. 

We show that in both classical and quantum cases, there is subdiffusion  which is
explained in the framework the CTRW. We also concern with a mechanism of subdiffusion,
which is due to the transitions between the 
strongly overlapped AMs.
This overlapping being a common feature of both classical and
quantum dynamics, leads to the clustering with an effective
trapping of the wave packet inside each cluster.
Therefore, the dynamics of the wave packet corresponds to the
CTRW, where the basic mechanism of subdiffusion is an entrapping
of the wave packet with delay, or waiting, times
distributed by the power law $w(t)\sim 1/t^{1+\alpha}$, where $\alpha$ is
the transport exponent. We show that $\alpha=1/3$ for the classical NLSE
and $\alpha=1/2$ for the quantum NLSE.

\section{Phase space dynamics and the Liouville operator}\label{sec:PSLO}
\def\theequation{\thesection.\arabic{equation}}
\setcounter{equation}{0}

First, we obtain the linear Liouville equation for $|\psi(x,t)|^2$ \cite{Iomin,iom1}.
Projecting Eq. (\ref{am1}) on the basis of the AMs
\be\label{am2}
 \psi(x,t)=\sum_{\omega_k}C_{\omega_k}(t)
\Psi_{\omega_k}(x)\equiv\sum_kC_k(t)\Psi_k(x)\, ,
 \ee %
we obtain a system of equations for coefficients of the expansion
$C_k$
\be\label{am3}
 i\prt_t{C}_k=\omega_kC_k+
\beta\sum_{k_1,k_2,k_3}A_{k_2,k_3}^{k,k_1}
C_{k_1}^*C_{k_2}C_{k_3}\, .
 \ee %
Here $A({\bf k})\equiv A_{k_2,k_3}^{k,k_1}$  is an overlapping
integral of the four AMs:
\be\label{am4}
 A_{k_2,k_3}^{k,k_1}
=\int\Psi_k(x)\Psi_{k_1}(x)\Psi_{k_2}(x)\Psi_{k_3}(x)dx\, .
 \ee%
The initial conditions for the system of Eqs. (\ref{am3}) are such
that $ \psi_0(x)=\sum_ka_k\Psi_k(x)=a_{l_0}\Psi_{l_0}$. Equations (\ref{am3})
correspond to a system of interacting nonlinear oscillators with
the Hamiltonian
\be\label{am6}
 H=\sum_k\omega_kC_k^*C_k+(\beta/2)\sum_{\bf
k}A_{k_2,k_3}^{k_1,k_4}
 C_{k_1}^*C_{k_4}^*C_{k_2}C_{k_3}\, .
\ee %
Therefore, Eqs. (\ref{am3}) are produced by the Poisson brackets
$\PBr$ by means of the Liouville operator
\be\label{am7}
 \hL=\frac{1}{i}\PBr  
= \frac{1}{i}\sum_k\left(\frac{\prt H}{\prt
C_k^*}\cdot\frac{\prt}{\prt C_k}-\frac{\prt H}{\prt
C_k}\cdot\frac{\prt}{\prt C_k^*}\right)\, .
 \ee %
Since $\hL H=0$ and $H(\{C,C^*\})=H(\{a,a^*\})$, we obtain that
the Liouville operator is an operator function of the initial
values:
\be\label{am7a}%
\hL=\frac{1}{i}\sum_k\left[\frac{\prt H}{\prt
a_k^*}\cdot\frac{\prt}{\prt a_k} -\frac{\prt H}{\prt
a_k}\cdot\frac{\prt}{\prt a_k^*}\right]
\ee%
and corresponds to an infinite system of linear equations $\prt_t{\bf
C}=\hL{\bf C}$, where ${\bf C}={\bf C}(\{a_k,a_k^*\})=(\dots ,
C_{k-1},C_k,C_{k+1},\dots)$ is an infinite vector. Thus, the
Liouville operator reads\footnote{For the evolution of the dynamical variables
like $C_k$ and $C_k^*$, it is a so called Koopman operator \cite{gaspard}.}
\be\label{am7b}%
\hL=-i\sum_k\omega_k\left(a_k\frac{\prt}{\prt a_k}-{\rm c.c.}\right)
-i\frac{\beta}{2}\sum_{\bf k} A_{k_2,k_3}^{k_1,k_4}
\left[a_{k_1}^*a_{k_2}a_{k_3}\frac{\prt}{\prt a_{k_4}} - {\rm c.c.}\right]\, ,   %
\ee%
where c.c. denotes the complex conjugation.
Finally, we obtain that the
system of {\it nonlinear} ordinary differential equations
(\ref{am3}) is replaced by a system of {\it linear} partial
differential equations:
\be\label{am8} %
 \prt_t{C}_k(t)=\hL C_k(t)\, , ~~~k= 1,2,\dots ,\, .
 \ee %

\section{Initial time dynamics: Perturbation approach}\label{sec:perturbation}
\def\theequation{\thesection.\arabic{equation}}
\setcounter{equation}{0}

Let us first understand how the Liouville operator (\ref{am7b}) describes the
dynamics of the physical variables. To this end, the Liouville operator is taken
as a combination of linear and nonlinear parts \cite{iom1}

\be\label{am7c8}%
\hL=-i(\hL_l+\beta\hL_{nl})\, ,  
\ee%
A formal solution of Eq. (\ref{am8}) is the expansion
\be\label{am10_a} %
\bar{C}_k(t)=\sum_{n=0}^{\infty}\Big[\frac{t^n}{n!} \hL
^na_k\Big]_{a_k=\delta_{k,l_0}}\,
.\ee %
The nonzero contribution to the first power over $t$ of the
expansion (\ref{am10_a}) is due to the term
\be\label{am10_b} %
\hL_{nl}^{(0)}=\frac{1}{2}\sum_kA(l_0,l_0,l_0,k)|a_{l_0}|^2\left(a_{l_0}\frac{\prt}{\prt
a_k}-c.c.\right)\, , %
\ee %
while $(\hL_{nl}-\hL_{nl}^{(0)})a_k\equiv 0$ is due to the initial
conditions $a_k=a_{l_0}\delta_{k,l_0}$. Moreover, the contribution of
$\hL_{nl}-\hL_{nl}^{(0)}$ without $\hL_{nl}^{(0)}$ is zero in all powers of
the expansion (\ref{am10_a}). For example, the $n$th power term
for $k\neq l_0$ is
\[
\left[\sum_{l\neq
l_0}A(l_0,l_0,l,l)|a_{l_0}|^2\prt_{\phi_l}\right]^{n}a_k
=i^nA^n(l_0,l_0,k,k)\delta_{k,l_0}\, .
\] %
It has a non zero contribution only in the power of the $n+1$
order after the action of the $\hL_{nl}^{(0)}$ term.
Therefore, keeping only the $\hL_{nl}^{(0)}$ term in Eq. (\ref{am10})
means neglecting $O(\beta^2t^2)$ terms in the expansion
(\ref{am10_a}). This solution is valid up to a time scale
$t<t_{\beta}=1/\beta$.

To obtain a solution in the framework of this approximation, first
we eliminate the linear term $\hL_l$ from Eq. (\ref{am8}) by
substitution
\be\label{am9}
 \bar{C}_k(t)=\exp(-i\hL_l t)C_k(t)\, .
  \ee %
After this substitution, Eq. (\ref{am8}) reads
\be\label{am10}
 \prt_t{\bar{C}}_k=
-i\beta\hL_{nl}(t)\bar{C}_k,~~~\hL_{nl}(t)=e^{-i\hL_lt}\hL_{nl}e^{i\hL_lt}\, .
 \ee%
 Taking into account that
\[\exp[-i\hL_lt]=\exp\left[-\sum_k\omega_kt\frac{\prt}{\prt\phi_k}\right]\]
is the phase shift operator for the complex values
$a_k=|a_k|e^{i\phi_k}$, we obtain
 \be   \label{am11}%
 \hL_{nl}(t)=\frac{1}{2}\sum_{\bf k} A({\bf k})\left[       
\exp[i\Delta\omega t] a_{k_1}^*a_{k_2}a_{k_3}\frac{\prt}{\prt
a_{k_4}} -{\rm c.c.}\right]\, ,
 \ee %
where
$\Delta\omega=\omega_{k_1}+\omega_{k_4}-\omega_{k_2}-\omega_{k_3}$.

Solutions of Eq. (\ref{am10}) for $k\neq l_0$ are functions which
are zero at $t=0$. These are
\be\label{am12b} %
\bar{C}_{k}(t)=a_k+\frac{\beta
A_1|a_{l_0}|^2a_{l_0}}{\Delta\omega+\beta A_0|a_{l_0}|^2}
\cdot \left(e^{-i\beta A_0|a_{l_0}|^2t}-e^{i\Delta\omega t}\right)\, .
 \ee %
Here $A_0=A(l_0,l_0,l_0,l_0)$ and $A_1\equiv
A_1(k)=A(l_0,l_0,l_0,k)/2\, , ~ k\neq l_0$, while $\Delta\omega$ now
is $\Delta\omega=\omega_k-\omega_{l_0}$. The complex conjugation
of Eq. (\ref{am12b}) can be a solution as well. A solution for $k=l_0$
is a function of $\phi_{l_0}-\beta A_0|a_{l_0}|^2t$, which
corresponds to the initial conditions. It reads
 \be\label{am12a}
\bar{C}_{l_0}(t)=a_{l_0}\exp(-i\beta A_0|a_{l_0}|^2t) \, .
\ee%
Using these analytical form for the coefficients
$\bar{C}_{k}(t)$ and Eq. (\ref{am9}), one obtains the solution of
the of NLSE (\ref{am1}) in the first order approximation over
$\beta$ as a sum
\bea\label{am13} %
&\psi(t)= a_{l_0}\exp(-i\omega_{l_1}t)\Psi_{l_0}(x)-
 4\beta |a_{l_0}|^2a_{l_0}\nonumber \\
 &\times\sum_k{}^{\prime}A_1(k)\frac{\sin
\left[\frac{\omega_k-\omega_{l_2}}{2}t\right]}{\omega_k-\omega_{l_2}}
 \sin\left[\frac{\omega_k+\omega_{l_2}}{2}t\right]
\Psi_k(x) \, , %
 \eea %
where $\omega_{l_1}=\omega_{l_0}+ \beta A_0|a|^2$ and
$\omega_{l_2}=\omega_{l_0}-\beta A_0|a|^2$, while prime means that
$k\neq l_0$. When $\beta\rightarrow 0$, we have at the
asymptotically large times $t_{\beta}\rightarrow\infty$ that
$\omega_{l_1}=\omega_{l_2}=\omega_{l_0}$, and the ${\rm sinc}$
function is
\[
\lim_{t\to\infty}
\frac{2\sin\left[\frac{\omega_k-\omega_{l_2}}{2}t\right]}
{\omega_k-\omega_{l_2}}
=2\pi\delta(\omega_k-\omega_{l_2})\, .\]%
The sum in Eq. (\ref{am13}) equals zero. Therefore, for $\beta=0$,
one obtains $\psi(t)=e^{-i\omega_{l_0}}\Psi_{l_0}(x)$ that
corresponds to a solution of the linear problem.

For nonzero values $\beta$ and $t<t_{\beta}$ the ${\rm sinc}$
function can be approximated by $t_{\beta}$ for
$\omega_k\approx\omega_{l_2}$. Then summation in Eq. (\ref{am13})
can be estimated by adding and subtracting the term with $k=l_0$.
Using the definition of the overlapping integrals $A_1(k)$ and
$\sum_k\Psi_k(x)\Psi_k(y)=\delta(y-x)$, one obtains an
approximation for Eq. (\ref{am13})
\be\label{am14} %
\psi(t)\sim \Psi_{l_0}(x)e^{-i\omega_{l_1}t}- 4\beta t
[\Psi_{l_0}^3(x)-A_0\Psi_{l_0}(x)]\sin(\omega_{l_2}t)
\, . %
\ee %
It means that at $t<t_{\beta}$ the wave function is localized and
its evolution corresponds to the periodic oscillations with the
frequencies $\omega_{l_1}$ and $\omega_{l_2}$. It is worth
mentioning that Eq. (\ref{am13}) is valid for both finite and
infinite systems (either discrete or continuous).

Consideration of the dynamics beyond $t>t_{\beta}$ relates to the
calculation of nonzero contributions of operators
$\Big[\hL_{nl}-\hL_{nl}^{(0)}\Big]^q$ and $\Big[\hL_{nl}^{(0)}\Big]^p$,
acting on the initial conditions. This combinatorics leads to
essential difficulties for analytical treatment. To overcome this
obstacle the dynamics of the initially localized states can be
considered qualitatively in the framework of a phenomenological
probabilistic approach.

\section{Liouville Equation}\label{sec:LE}
\def\theequation{\thesection.\arabic{equation}}
\setcounter{equation}{0}

The Liouville equation
is valid for any functions of the initial conditions
$\{a_k,~a_k^*\}$. In particular, introducing the function
$F_{k,k'}(t)=C_k(t)\cdot C_{k'}^*(t)$, one has the Liouville
equation:
\[\prt_tF_{k,k'}(t)=\hL F_{k,k'}(t)\, ,
~~~F_{k,k'}(t=0)=F_{k,k'}^{(0)}=a_ka_{k'}^*\, .\]
Therefore, the probability amplitude
\[\clP(x,t)=|\psi|^2=\sum_{k,k'}F_{k,k'}(t)\Psi_k(x)\Psi_{k'}(x)\, ,\]
as a function of the initial conditions,  satisfies the Liouville
equation as well:
\be\label{trueWF}   %
\prt_t\clP=\hL\clP\, .
\ee %
Here the initial condition is
\be\label{icond} %
\clP(x,t=0)=\clP_0(x)=\sum_{k,k'}F_{k,k'}^{(0)}\Psi_k(x)\Psi_{k'}(x)\, .  %
\ee  %
Eventually, one considers the dynamics of the probability distribution 
function (PDF) $\clP$ in the framework of
Liouville equation (\ref{trueWF}), which is the \textit{linear}
equation with a formal solution in the exponential form
\be\label{am10_abc} %
\clP(x,t)=e^{\hL t}\clP_0(x)=
\sum_{k,k'}\Psi_k(x)\Psi_{k'}(x)\sum_{n=0}^{\infty}\Big[\frac{t^n}{n!}
\hL ^n\Big]F_{k,k'}^{(0)}\, .  %
\ee %
It is worth noting that the linear property of the Liouville
equation (\ref{trueWF}) and its formal solution (\ref{am10_abc}) are
important for the probabilistic approach.

According to the values of the overlapping integrals (\ref{am4}), we divide
the transitions between the localized states into two main groups.
The first one corresponds to the exponentially small overlapping
integrals and the second one corresponds to the strong overlapping
between four AMs when the overlapping integrals are of the order
of $1$.  In the case of strong overlapping, the AMs form clusters,
where the wave functions have the same averaged coordinates for
each cluster. Consequently, all transitions inside one cluster do
not lead to any appreciable differences in the coordinate space,
and we regard these transitions as \textit{trapping} of the wave
packet, or a particle, inside this cluster. Contrary to that,
transitions due to the exponentially small overlapping integrals
between the AMs belonging to different clusters lead to a change
of the space coordinates that can be accounted for. We call these
processes \textit{jumps}. Contributions of trappings and jumps to
the wave packet spreading described by Eqs. (\ref{trueWF}) and
(\ref{am10_a}) are different, and correspond to different
probabilistic interpretations.

Therefore, the Liouville operator
can be considered as a sum of two operators $\hL=\hL_0+\hL_1$, where
$\hL_0$ corresponds to the dynamics inside a cluster, determined by the
overlapping integrals of the order of 1, while $\hL_1$ corresponds to jumps
between these clusters due to the exponentially small overlapping integrals.
Thus considering this dynamics in the ``interaction picture'', where $\hat{L}_1(t)=e^{-\hat{L}_0t}\hat{L}_1e^{\hat{L}_0t}$, one obtains
the solution of the Liouville Eq. (\ref{ctqw_1}) in the iterative form
\begin{eqnarray}\label{MEa}
\clP(x,t)&=&e^{\hat{L}_0t}\left[1+\int_0^tdt_1\hat{L}_1(t_1)+
\int_0^tdt_1\int_0^{t_1}dt_2\hat{L}_1(t_1)\hat{L}_1(t_2)\right. \nonumber \\
&+& \left.
\dots +\int_0^rdt_1\dots \int_0^{t_{n-1}}dt_n\hat{L}_1(t_1)
\dots\hat{L}_1(t_n)+\dots \right] \nonumber \\
&=& e^{\hL_0t}\clP_0(x)+\int_0^tdt'e^{(t-t')\hL_0}\hL_1\clP_0(x,t')\, .
\end{eqnarray}
The probabilistic interpretation of the last expression is as
follows. The first term in Eq. (\ref{MEa}) corresponds to a
particle, which at the initial time is at the position $x$ and there
are no jumps to another clusters until time $t$, and therefore
$e^{\hL_0t}\clP_0(x)\rightarrow W(t)\clP_0(x)$. Here $W(t)$
denotes the probability of no jump until time $t$, since the particle is
inside the same cluster. It is described by all possible transitions
between AMs inside one cluster, which is
characterized by the coordinate $x$. The last term
in Eq. (\ref{MEa}) corresponds to a particle to be at position $x$ at
time $t$ due to all possible jumps from points $(x',t')$ with the
transition probability $\clG(x-x';t-t')$ between different clusters,
which is determined by operator $\hL_1$. It is a composite operator,
which consists of both inter-cluster dynamics due to the operator $\hL_1$
and trapping inside one cluster dynamics due to the operator $\hL_0$.
Note, also, that when $\beta=0$, the transition
probability is zero: $\clG(x-x';t-t')=0$. Therefore, Eq.
(\ref{MEa}) now reads
\be\label{mastereq}  %
\clP(x,t)=W(t)\clP_0(x)+
\int_0^tdt'\int_{-\infty}^{\infty}dx'\clG(x-x';t-t')\clP(x',t')\,
.\ee  %

This linear property can be used now for the continuous time
random walks (CTRW) approach to obtain the Montroll-Weiss equation
\cite{klafter,shlesinger,montweiss}. We also express here the
dependence of the transition probability on $\beta$ in the
explicit form to stress that, for $\beta\rightarrow 0$, the
dynamics is localized. This dependence on $\beta$ will be
reflected in a generalized transport coefficient. It should be
admitted that CTRW processes are connected with a continuous time
generalization of the Chapman-Kolmogorov equation
\cite{metzler2000,zaslavsky1994}.

\section{CTRW}\label{sec:CTRW}
\def\theequation{\thesection.\arabic{equation}}
\setcounter{equation}{0}

In what follows we consider the dynamics of the initial wave
packet $\clP_0(x)$ in the framework of the probabilistic approach,
where the dynamics of the wave packet is considered as the CTRW.
Since the dynamics of the probability distribution function (PDF)
$\clP(x,t)$ is governed by the same
Liouville operator in Eqs. (\ref{trueWF}) and (\ref{am10_a}), the
overlapping integrals $A({\bf k})=A_{\bf k}$, defined in Eq. (\ref{am4}),
play the dominant role in the
wave packet spreading. As follows from Eq. (\ref{am10_a}), the
overlapping integrals determine the spread of the initially
localized wave packet $\clP_0(x)$ over all the AMs as
transitions from one mode to another. Since all states are
localized, these transitions between states determine the
transitions (or jumps) in the coordinate space as well.

As discussed, transitions between strongly overlapped
AMs contribute to the trapping of the wave packet
inside one cluster, while transitions between
AMs belonging to different clusters contribute to jumps.
It is reasonable to assume that the transitions
between these different states are independent of each other; therefore, the jumps are
independent and obey the Markov property, and
the PDF of a jump $p(x)$ is determined by the overlapping integrals as
$p(x)=\xi\exp(-\xi|x|)/2$, and $\xi=1/R$ is an inverse
localization length of the eigenstates $\Psi_k(x)$ in Eq. (\ref{am2}).

The trapping is associated with clusters with effective lengths
$\Delta$. Due to the exponential localization, these values are
distributed by the exponential law $P_{\rm cl}\left(\Delta
\right)=\Delta_0^{-1}\exp\left(-\Delta /\Delta_0\right)$, where
$\Delta_0$ is the effective (maximal) length of a cluster. The effective lengths
are determined by overlapping integrals of four wave functions;
therefore, the minimum
length of the cluster is $\Delta =R$, while the effective length is
$\Delta_0 =3R$, which is a maximal distance between four strongly overlapping AMs.
Now the probability that a particle exits this cluster and jumps to
another one is of the order of $\sim \exp\left(-\Delta /R\right)$.
This value is also proportional to the inverse waiting time,
$t\sim\exp\left(\Delta /R\right)$. The probability to find the
waiting time in the interval $(t,~t+dt)$ is equal to the
probability to find the corresponding trapping length in the
interval $(\Delta,~\Delta+d\Delta)$, namely, $w(t)dt=P_{\rm
cl}\left(\Delta \right)d \Delta $. Therefore, after simple
calculations one obtains that the PDF of the waiting times is
\be\label{pdf_a} %
w(t)=P_{\rm cl}\left(\Delta \right)\frac{d \Delta }{dt} \sim
\frac{1}{(t/\tau)^{1+\alpha}}\, ,       %
\ee    %
where $\alpha=R/\Delta_0=1/3$ and $\tau$ is a time scale related to
the trapping\footnote{For example, taking $\Delta=\Delta_0$, one obtains
$t=\tau=e^{-1/3}$.}. It follows that the mean waiting time is
infinite. Taking into account that the waiting time PDF is
normalized, we have
\be\label{pdf_w} %
w(t)=\frac{w_0}{1+(t/\tau)^{4/3}}\, ,    %
\ee %
such that $ \int_0^{\infty}w(t)dt =1$, while
$\int_0^{\infty}tw(t)dt=\infty$, were
$w_0$ is a normalization constant.
We also admit that according the definition
\begin{equation}\label{w-W}
W(t)=\int_t^{\infty}w(t')dt'\, ,
\end{equation}
where $W(t)=1-\int_0^tw(t')dt'$ denotes the probability of no jump
during the time interval $(0,t)$, introduced in Eq.(\ref{mastereq}).
Performing the Fourier transform
$\bar{p}(k)=\hat{\clF}[p(x)]$ and the Laplace transform
$\tilde{w}(u)=\hat{\clL}w(t)$, we obtain the Montroll-Weiss equation
\cite{montweiss} from Eq. (\ref{mastereq})
\be\label{MW}  %
\bar{\tilde{\clP}}(k,u)=\hat{\clF}\hat{\clL}\clP=
\frac{1-\tilde{w}(u)}{u}\cdot\frac{\bar{\clP}_0(k)}{1-\bar{p}(k)\tilde{w}(u)}\,
, \ee  %
where the transition probability $\clG(x;t)$ is considered in the multiplicative form
$\bar{\tilde{\clG}}(k,u)=\bar{p}(k)\tilde{w}(u)$.
Equation (\ref{MW}) can be simplified for the long time $u\ll 1$ and
the large scale $k\ll 1$ asymtotics that corresponds to the
diffusion limit $(k,u)\rightarrow(0,0)$. Taking into account the
Fourier $\bar{p}(k)$ and the Laplace $\tilde{w}(u)$ images in Eq.
(\ref{MW}):
\bea\label{asympt}  %
\bar{p}(k)&=&\frac{1}{1+\beta R^2k^2}\approx 1-\beta R^2k^2\,  , \nonumber \\
\tilde{w}(u)&=&\frac{1}{1+(u\tau)^{\alpha}}\approx  1-(u\tau)^{1/3}\, ,
\eea  %
we obtain for the
PDF in the Fourier-Laplace domain (see also \cite{klafter})
\be\label{MWasy}      %
\bar{\tilde{\clP}}(k,u)=
\frac{\bar{\clP}_0(k)/u}{1+D_{\frac{1}{3}}u^{-1/3}k^2}\, ,
\ee %
where $D_{\frac{1}{3}}=\beta R^2/\tau^{1/3}$ is a generalized diffusion
coefficient\footnote{The nonlinear parameter
is explicitly introduced in the generalized diffusion coefficient,
such that for $\beta=0$, there is no
any transport and the initial wave packed remains localized.}. 
Using the Laplace transform of the fractional integration
\[\hat{\clL}\left[\prt_t^{-\alpha}f(t)\right]= \hat{\clL}\left[
\frac{1}{\Gamma(\alpha)} \int_0^t\frac{f(\tau)d\tau}{(t-\tau)^{1-\alpha}}\right] =
u^{-\alpha}f(u)\,  ,~~~\alpha>0\, ,\]
one  obtains the fractional integral equation
\be\label{fieq}
\clP(x,t)-\clP_0(x)=\prt_t^{-\alpha}D_{\alpha}\prt_x^2\clP(x,t)\, .
\ee
Differentiating this equation with respect to time, one obtains
that the CTRW is described by the fractional Fokker-Planck
equation\footnote{The solution of the FFPE is obtained in the form of the Fox
$H$ function, presented in Appendix B (see \textit{e.g.}, \cite{klafter})
and its asymptotic behavior corresponds to the stretched exponential
function $\clP(y)\sim \frac{1}{\sqrt{D_{\alpha}t^{\alpha}}}
y^{-(1-\alpha)/(2-\alpha)}e^{-y^{2/(2-\alpha)}}$,
where $y\equiv\frac{|x|}{\sqrt{D_{\alpha}t^{\alpha}}}\gg 1$. For
$\alpha=1$ it corresponds to the normal Gaussian distribution.}
(FFPE)
\be\label{ffpe} %
\prt_t\clP(x,t)-D_{\alpha}\prt_t^{1-\alpha}\prt_x^2\clP(x,t)=0\, ,
\ee %
where $\prt_t^{\alpha}$ is a designation of the Riemann-Liouville fractional
derivative
\[\prt_t^{\alpha}f(t)=\frac{d}{dt}\prt_t^{\alpha-1}f(t)=
\frac{1}{\Gamma(1-\alpha)}\frac{d}{dt}
\int_0^t\frac{f(\tau)d\tau}{(t-\tau)^{\alpha}}\,  \] %
with $\alpha=1/3$, while in general case $0<\alpha<1$.
From Eq. (\ref{ffpe}) one obtains for the MSD $\lgl
x^2(t)\rgl=\int_{-\infty}^{\infty}x^2\clP(x,t)dx$ the differential
equation:
\[\frac{d}{dt}\lgl x^2(t)\rgl =
\frac{2D_{\alpha}t^{\alpha-1}}{\Gamma(\alpha)}\, .\] %
Here $\Gamma(z)$
is the gamma function, $x(t=0)=0$, and we use the following
property of the fractional derivative
$\prt_t^{\alpha}1=t^{-\alpha}/\Gamma(1-\alpha)$. Therefore, Eq.
(\ref{ffpe}) describes subdiffusion
\be\label{subdif} %
\lgl x^2(t)\rgl=\frac{2D_{\alpha}t^{\alpha}}{\Gamma(1+\alpha)}=
\frac{2D_{\frac{1}{3}}t^{1/3}}{\Gamma(4/3)}\,
, \ee %
with the transport exponent $\alpha=1/3$. This result exactly corresponds
to the numerical \cite{fks,fks1,Skokos2010} and analytical
\cite{MI2012,MI2015,MI2014} studies.

\section{Quantum NLSE}\label{sec:QNLSE}
\def\theequation{\thesection.\arabic{equation}}
\setcounter{equation}{0}

Now we concern with a quantum counterpart of the NLSE
(\ref{am1}). In this case, an initial wave packet spreading is
governed by the quantum nonlinear Schr\"odinger
equation, which leads to the
quantum counterpart of the Hamiltonian (\ref{am6})
\begin{equation}\label{q-cl}
\hat{H}=\sum_k\hh\omega_k\hC_k^{\dag}\hC_k+\frac{\hh^2\beta}{2}\sum_{\mathbf{k}}
A_{\mathbf{k}}\hC_{k_1}^{\dag}\hC_{k_2}^{\dag}\hC_{k_3}\hC_{k_4}\, ,
\end{equation}
where $\hh$ is a dimensionless Planck constant and the operators are not commute
\begin{equation}\label{q-commute}
[\hC_i,\hC_j^{\dag}]=\delta_{ij}\, .
\end{equation}
The linear frequency is shifted by the nonlinear term due to the commutation
$\omega_k\rightarrow \omega_k+\hh\beta A_0$, where $A_0=A_{kk}^{kk}$ is the diagonal
overlapping integral in Eq. (\ref{am4}) and $A_{\mathbf{k}}\equiv A_{k_3,k_4}^{k_1,k_2}$.
One should recognize that the classical NLSE (\ref{am1}) is independent of the
Planck constant. Therefore, for the quantum analysis, an effective dimensionless
Planck constant $\hh$ is introduced explicitly.

For the quantum mechanical analysis
we use a technique of mapping the Heisenberg equation of motion on a basis
of the coherent states \cite{st1982,biz1981,bikt1986}. At the initial moment
$t=0$, one introduces the coherent states vector $|\mathbf{a}\rgl=\prod_q|a_q\rgl$
as the product of eigenfuctions
of annihilation operators $\ha_q=\hC_q(t=0)$, such that $\ha_q|a_q\rgl=a_q|a_q\rgl$
and correspondingly $\ha_q|\mathbf{a}\rgl=a_q|\mathbf{a}\rgl$, where also the coherent state is constructed from a vacuum state $|0\rgl$
\begin{equation}\label{q2}
|a_q\rgl=\exp[a_q\ha_q^{\dag}-a_q^{*}\ha_q]|0\rgl\, , ~~~~\ha_q|0\rgl=0\, .
\end{equation}
Introducing $c$-functions
\begin{equation}\label{q3}
\clC_q(t)=\lgl\mathbf{a}|\hC_q(t)|\mathbf{a}\rgl=
\clC_q(t|\mathbf{a}^{*},\mathbf{a})\, ,
\end{equation}
one maps the Heisenberg equation of motion on the basis $|\mathbf{a}\rgl$
as follows
\begin{equation}\label{q4}
i\hh\dot{\clC}_q(t)=\lgl\mathbf{a}|\left[\hC_q(t)\hat{H}-
\hat{H}\hC_q(t)\right]|\mathbf{a}\rgl\,.
\end{equation}
Accounting Eqs. (\ref{q2}) and (\ref{q3}), one obtains the mapping rules
\begin{eqnarray}\label{q5}
\lgl\mathbf{a}|\hC_q(t)\ha_q^{\dag}|\mathbf{a}\rgl&=&
e^{-|a_q|^2}\frac{\prt}{\prt a_q}e^{|a_q|^2}\clC_q(t)\, , \nonumber \\
\lgl\mathbf{a}|\ha_q\hC_q(t)|\mathbf{a}\rgl&=&
e^{-|a_q|^2}\frac{\prt}{\prt a_q^{*}}e^{|a_q|^2}\clC_q(t)\, .
\end{eqnarray}
The Hamiltonian is the integral of motion
$\hat{H}(\{\hC_q^{\dag},\hC_q\})=\hat{H}(\{\ha_q^{\dag},\ha_q\})$, therefore the mapping rules
(\ref{q5}) yield equation of motion (\ref{q4}) for $\clC_q(t)$ in the closed form
\begin{equation}\label{q6}
i\dot{\clC}_q(t)=\hat{K}\clC_q(t)\, ,
\end{equation}
where
\begin{eqnarray}\label{q7}
&{}&\hat{K}=\frac{1}{\hh}e^{-\sum_k|a_k|^2}\left[
\hat{H}(\{\frac{\prt}{\prt a_q}\,,a_q\})
-\hat{H}(\{a_q^{*}\,,\frac{\prt}{\prt a_q^{*}}\,,\})\right]e^{\sum_k|a_k|^2} = \nonumber \\
&{}&\sum_q\left[\omega_q a_q\frac{\prt}{\prt a_q}-c.c\right]
+\frac{\hh\beta}{2} \sum_{\mathbf{q}}A_{\mathbf{q}}
\left[2a_{q_1}a_{q_2}a_{q_3}^{*}\frac{\prt}{\prt a_{q_4}}
+ a_{q_1}a_{q_2}\frac{\prt}{\prt a_{q_3}}\frac{\prt}{\prt a_{q_4}}
 -c.c\right] \, . \nonumber \\
 &{}&
\end{eqnarray}
In the limit $\hh\rightarrow 0$, the second derivative terms vanish,
$\hh a_{q_1}a_{q_2}\frac{\prt}{\prt a_{q_3}}\frac{\prt}{\prt a_{q_4}}
\rightarrow 0$, and the operator (\ref{q7}) reduces to the classical
Liouville operator \cite{Iomin} in Eq. (\ref{am7b}.

Following the strategy of the
construction of the kinetic equation (\ref{ffpe}) for
the classical NLSE, we consider the density operator $\hat{\rho}=|\hat{\psi}|^2$ and map it
on the basis of the coherent states $|\mathbf{a}\rgl$ in Eq. (\ref{q2}), such that
\begin{equation}\label{q-density-rho}
\clP_Q(t)\equiv\clP_Q(\mathbf{a}^*,\mathbf{a},t)=
\lgl\mathbf{a}|\hat{\rho}(t)|\mathbf{a}\rgl\, .
\end{equation}
Therefore, from the mapping rules one obtains the Liouville equation
for the mean probability density
\begin{equation}\label{ctqw_1}
\partial_t\clP_Q(t)=\hL\clP_Q(t)\, .
\end{equation}
Here the quantum ``Liouville'' operator is determined in Eq. (\ref{q7})
\begin{equation}\label{K-L}
\hL=-i\hat{K}\, .
\end{equation}
In the complete analogy with Eq. (\ref{trueWF}), the Liouville operator
in Eq. (\ref{K-L}) is considered as a sum of two operators $\hL=\hL_0+\hL_1$, where
$\hL_0$ corresponds to the dynamics inside clusters, determined by the
overlapping integrals of the order of $1$, while $\hL_1$ corresponds to jumps
between these clusters due to the exponentially small overlapping integrals.
Eventually, one arrives at the master equation (\ref{mastereq}) for the quantum
mean probability density $\clP_Q(t)$, which coincides exactly with
Eq. (\ref{mastereq})
\begin{equation}\label{qmastereq}  %
\clP_Q(x,t)=W_Q(t)\clP_0(x)+
\int_0^tdt'\int_{-\infty}^{\infty}dx'\clG_Q(x-x';t-t')\clP_Q(x',t')\,
.\end{equation}

In this construction of the quantum master equation, as a quantum CTRW,
we follow a van-Kampen coarse-graining
procedure (see details of the discussion in \cite{kreuzer}) by lumping a set of
quantum states making a trapping cluster, which is in
complete analogy with the classical CTRW, constructed in \cite{Iomin}.
The transition probabilities in Eq. (\ref{qmastereq})
(Pauli-van-Kampen master equation) reduce to their classical analogs of the
transition probabilities, since the quantum ``Liouville'' operator determined in Eqs.
(\ref{q7}) and (\ref{K-L}) contains the same overlapping integrals as its classical
counterpart with $\hh=0$. As the result,
the coarse-grained process of quantum transitions between the trapping
clusters is also a Markov process. Transitions between quantum
states inside the cluster do not contribute to the transport (to the spreading of the
initial wave packet in the $x$ coordinate), while transitions between any
states of the different trapping clusters lead to the spreading of the wave
packet in the $x$ coordinate.
The main contribution to this transport in the chain of the localized
AMs (in a more general case, Wannier states) is due to the transitions between the
nearest neighbor clusters and this is determined by the jump length PDF $p_Q(x)$.
The quantum transitions between coherent states inside the cluster do not contribute
to the transport with trapping, or delay times, distributed by
the waiting time PDF $w_Q(t)$. The latter is determined by
quantum decay processes, therefore the probability
to find a quantum particle inside the trapping cluster after time $t$ (or no jump
during time interval $(0,t)$) is
\begin{equation}\label{W-w}
W_Q(t)=\int_t^{\infty}w_Q(t')dt\, .
\end{equation}
Following the standard CTRW approach of Sec.~\ref{sec:CTRW}
\cite{klafter,shlesinger,montweiss},
the composite transition
probability operator $\clG_Q(x-x';t-t')$ in Eq. (\ref{qmastereq}) is considered in the multiplicative form by analogy with the classical CTRW.
Performing the Fourier transform $\bar{p}_Q(k)=\hat{\clF}[p_Q(x)]$
and the Laplace transform $\tilde{w}_Q(u)=\hat{\clL}[w_Q(t)]$, we obtain that
the quantum distribution function is governed by the Montroll-Weiss equation in
the Fourier-Laplace space $\bar{\tilde{\clP}}_Q(k,u)=\hat{\clL}\hat{\clF}[\clP_Q(x,t)]$
\begin{equation}\label{ctqw_2}
\bar{\tilde{\clP}}_Q(k,u)=\hat{\clF}\hat{\clL}\clP_Q=
\frac{1-\tilde{w}_Q(u)}{u}\cdot\frac{\bar{\clP}_0(k)}{1-\bar{p}_Q(k)\tilde{w}_Q(u)}\, .
\end{equation}
Here it was assumed that the transitions
between different clusters are classical processes, which are
independent of each other.  Therefore, the jumps are
independent and obey the Markov property, where
the PDF of a jump $p_Q(x)$ is
determined by the overlapping integrals as
$p_Q(x)=p(x)=\xi\exp(-\xi|x|)/2$, and $\xi=1/R$ is an inverse
localization length of the AM $\Psi_k(x)$ in Eq. (\ref{am2}).

The situation with waiting time PDF $w_Q(t)$ changes drastically.
The transitions inside a cluster are pure quantum transitions,
which determine the survival probability.  It corresponds to the inverse
amplitude of the population of a quantum states $|a_q\rgl$ with $q\neq k$,
described by $\clC_q(t)$ with the initial condition
$\clC_q(t=0)=a_k$.

Therefore, we study the dynamics of four modes, which is governed by the
Hamiltonian (\ref{q-cl}), where we take into account only resonant terms in the interaction.
These resonant processes are the fastest and
as it is shown in Refs.~\cite{MI2012,MI2015,MI2014}, these terms have the strongest
contribution to the wave packet spreading.
Therefore, locally these modes are described by the same overlapping integrals $\sim A_0\delta_{k_1+k_2-k_3-k_4,0}$, where, however, the interaction
$A_0$ varies for different clusters.

\section{Four-modes stability analysis}\label{sec:FMSA}
\def\theequation{\thesection.\arabic{equation}}
\setcounter{equation}{0}

The local Hamiltonian of the quantum NLSE in a cluster reads
\begin{equation}\label{q1}
\hat{H}_l=\sum_k\hh\omega_k\hC_k^{\dag}\hC_k+\frac{\hh^2\beta A_0}{2}\sum_{\mathbf{k}}
\hC_{k_1}^{\dag}\hC_{k_2}^{\dag}\hC_{k_3}\hC_{k_4}\delta_{k_1+k_2-k_3-k_4,0}\, ,
\end{equation}
In this case the $c$ number Heisenberg equation (\ref{q6})
\begin{equation}\label{q6a}
i\dot{\clC}_q(t)=\hat{K}\clC_q(t)\, ,
\end{equation}
is controlled by the local Liouville operator, which describes the quantum
dynamics inside one cluster.
\begin{eqnarray}\label{q7a}
&{}&\hat{K}
=\sum_q\left[\omega_q a_q\frac{\prt}{\prt a_q}-c.c\right]
+\frac{\hh\beta A_0}{2} \nonumber \\
&\times&\sum_{\mathbf{q}}
\left[2a_{q_1}a_{q_2}a_{q_3}^{*}\frac{\prt}{\prt a_{q_4}}
+ a_{q_1}a_{q_2}\frac{\prt}{\prt a_{q_3}}\frac{\prt}{\prt a_{q_4}}
 -c.c\right]\delta_{q_1+q_2-q_3-q_4,0} \, .
\end{eqnarray}
Taking the initial conditions in the form $\clC_q(t=0)=a_k$ for $q=k$ and
$\clC_q(t=0)=0 $ for $ q\neq k$, one obtains by the straightforward substitution
that the solution of Eq. (\ref{q6a}) is a periodic wave with the finite amplitude
\cite{bikt1986,biz1981}
\begin{eqnarray}\label{q8}
\clC_k(t)&=&a_k\exp(-i\omega_kt)\left[(e^{-i\hh\beta A_0 t}-1)|a_k|^2\right]\, ,
\nonumber \\
\clC_q(t)&=&0\, , ~~~q\neq k\, .
\end{eqnarray}

Now we follow the analysis of Ref. \cite{bikt1986} to study the stability
of this initially populated
state. To this end we study the dynamics of the quantum states
with $q\neq k$. We also study the
dynamics of the survival probability to stay inside the cluster.
Obviously, it corresponds to the inverse amplitude of the population
of the quantum states
$|a_q\rgl$ with $q\neq k$, which are described
by $\clC_q(t)$. The stability of the solution (\ref{q8}) can be studied in the framework
of the resonant four-wave decay processes $2k\rightarrow k\pm p$.
We take into account that due to the non-resonant interactions
all modes with $q\neq k$ are instantly populated at the initial time $t=0$,
such that the conditions for the initial amplitudes are $|a_q|\ll|a_k|$.
In this case, one looks for the solution of $\clC_{k+p}$  as
an expansion over the powers of $a_q$. Taking into account only the first orders
of the expansion for the ``small'' waves, we have
\begin{equation}\label{q9}
\clC_{k+p}(t|\{a_q^{*},a_q\})=f_0(t|a_k^{*},a_k)+\sum_{q\neq 0}
[f_q(t|a_k^{*},a_k)a_q+ \tilde{f}_q(t|a_k^{*},a_k)a_q^{*}]+{\rm o}(a_q,a_q^{*}) \, .
\end{equation}
Due to the initial condition $\clC_{k+p}(0)\equiv\clC_{k+p}(0|\{a_q^{*},a_q\})=a_{k+p}$,
one obtains the initial conditions for the amplitudes
$f_0(t) \equiv f_0(t|a_k^{*},a_k)$, $f_q(t)\equiv f_q(t|a_k^{*},a_k)$, and
$\tilde{f}_q(t)\equiv\tilde{f}_q(t|a_k^{*},a_k)$ as follows
\begin{eqnarray}\label{q10}
&f_0(0)=a_k\, , ~~ f_q(0)=\tilde{f}_q(0)=0\, ~~~& \mbox{for $p=0$}\, ,\nonumber \\
&f_0(0)=0\, , ~~  f_q(0)=\delta_{q,p}\, , ~~\tilde{f}_q(0)=0\, ~~~
& \mbox{for $p\neq 0$}\, .
\end{eqnarray}
The dynamics of the functions $f_q$ and $\tilde{f}_q$
determine the dynamics of small amplitudes $\clC_q(t)$.
Substituting expansion (\ref{q9}) in Eq. (\ref{q6a}) and taking into account
expression (\ref{q7a}) for the quantum operator and collecting the same powers of the small
initial amplitudes $a_{k\pm p}$ and $a_{k\mp p}^{*}$, one obtains the system of
equations \cite{bikt1986}
\begin{eqnarray}\label{q11}
i\frac{\prt f_0}{\prt t} &=& \hat{K}_0f_0 \nonumber \\
i\frac{\prt f_q}{\prt t} &=& \hat{K}_0f_q+[\omega_{k+q}+2\hh\beta A_0|a_k|^2]f_q
+2\hh\beta A_0a_k\frac{\prt}{\prt a_k}f_q-\hh\beta A_0a_k^{*2}\tilde{f}_{-q}\, , \nonumber \\
i\frac{\prt\tilde{f}_{-q}}{\prt t} &=& \hat{K}_0\tilde{f}_{-q}-
[\omega_{k-q}+2\hh\beta A_0|a_k|^2]\tilde{f}_{-q}
-2\hh\beta A_0a_k\frac{\prt}{\prt a_k}\tilde{f}_{-q}+\hh\beta A_0a_k^2 f_{q}\, . \nonumber \\
&{}&
\end{eqnarray}
Here
\begin{equation}\label{q12}
\hat{K}_0=[\omega_k  + \hh\beta A_0|a_k|^2]a_k \frac{\prt}{\prt a_k} +
\frac{\hh\beta A_0}{2}a_k^2\frac{\prt^2}{\prt a_k^2} -c.c. \, .
\end{equation}

Solution of the first equation in system (\ref{q11}) coincides with the nonzero solution
in Eq. (\ref{q8}), and in this first order approximation it does not describes the escape
rate, and we disregard this solution. In the next two equations the nonzero solutions
exist only for $q=\pm p$. Then performing the variable change
\begin{equation}\label{q13}
f_p=fe^{-i(2\omega_k-\omega_{k-p}+\hh\beta A_0)t}\, , ~~~\tilde{f}_{-p}=
\frac{\alpha_k}{\alpha_k^{*}}g\exp[-i(2\omega_k-\omega_{k-p}+\hh\beta A_0)t]\, ,
\end{equation}
and introducing the angle variable $I=\hh|a_k|^2$, as it is shown in appendix C,
one obtains equations for the amplitude $f=f(I,t)$ and $g=g(I,t)$
\begin{eqnarray}\label{q14}
i\prt_tf &=& -\Delta\omega f+2\hh\beta A_0 I \prt_If +2\beta A_0I f -\beta A_0 I g\, ,
\nonumber \\
i\prt_tg &=&  \beta A_0 I f\, .
\end{eqnarray}
Here we take into account that the resonant four-mode
decay takes place at both the momentum conservation $2k\rightarrow (k+p)+(k-p)$
and the energy detuning $\Delta\omega=2\omega_k-\omega_{k+p}-\omega_{k-p}-\hh\beta A_0$.
Introducing dimensionless variables $\mcT=\Delta\omega t$, $v=\beta A_0I/\Delta\omega$,
and $\vep=\hh\beta A_0/\Delta\omega$, one obtains from Eq. (\ref{q14})
\begin{equation}\label{q15}
i\prt_{\mcT}f=(2v-1)f- vg+2\vep y\prt_vf\, , ~~~~~
i\prt_{\mcT}g= v f\, .
\end{equation}
The initial conditions are $f(\mcT=0)=1$ and $g(\mcT=0)=0$.

Equation (\ref{q15}) describes the decay instability in the quantum case.
Following \cite{bikt1986}, they will be referred to as equations of quantum decay.

\subsection{Solution for the quantum four-mode decays}\label{sec:solution}

A careful mathematica analysis of the system (\ref{q15}) has been performed in
Ref.~\cite{bikt1986} in the form of the semiclassical expansion (see also recent
results \cite{bt2006,bt2008}). The system (\ref{q15}) is of mixed type with
hyperbolic degeneracy on the line $v = 0$. The general theory yields merely that
it has a real analytic solution in the three-dimensional space $(\tau,\, v,\, \vep)$
in some neighborhood of the plane $\mcT = 0$. As it is shown in
\cite{bikt1986,bt2006,bt2008}, the quantum decays run not faster than exponential $\exp(\sigma\mcT)$, where $\sigma$ does not depend on $\mcT$, such that this
property enables one to apply the Laplace transform in the analysis of
equations (\ref{q15}).

Applying the Laplace transform in time $f_u(v)=\hclL[f(\mcT)]$ and excluding
$g_u(v)$, one obtains an ordinary equation for $f_u(v)$
\begin{equation}\label{q16}
2\vep v\frac{d \, f_u}{d\,v} +(+iv/u-iu-2)f_u=-i\, , ~~~
~~f_u(v=0)=\frac{i}{2+iu}\, .
\end{equation}
The solution of equation (\ref{q15} is
\begin{equation}\label{q17}
f(v,\mcT)=\hclL^{-1}\left[-ie^{\frac{1}{2\vep}(\frac{v^2}{2iu}-2v)}
\sum_{n=0}^{\infty}\frac{1}{n!}\Big(\frac{v}{4\vep}\Big)^n\sum_{m=0}^n
\frac{C_m^n}{2\vep(m+n)-(iu+2)}\Big(-\frac{v}{4iu}\Big)^m\right]\, ,
\end{equation}
where $C_m^n$ are binominal coefficients. For the inverse Laplace transform
we expand the exponential in Eq. (\ref{q15}) in the series, which yields
(see Appendix D)
\begin{equation}\label{q18}
f(v,\mcT)=e^{-\frac{v}{\vep}}\sum_{n=0}^{\infty}\frac{1}{n!}\Big(\frac{v}{\vep}\Big)^n
\sum_{m=0}^{n}C_m^n\Big(\frac{v}{4i}\Big)^mF_{m,n}(\mcT)\, ,
\end{equation}
where the time dependent term $F_{m,n}(\mcT)$ is obtained in Appendix D in the form of
the Bessel functions
\begin{equation}\label{q19}
F_{m,n}(\mcT)=(ic_{m,n})^{-m}\left[e^{ic_{m,n}\mcT}e^{-\frac{v^2}{4\vep c_{n,m}}}
+\sum_{k=1}^{\infty}\frac{J_n\big(\Lambda\big)}{(-ic_{m,n})^k}+
\sum_{k=0}^{m-1}\Big(-\frac{4\vep c_{n,m}}{v^2}\Big)^{k}J_n\big(\Lambda\big)\right]\, ,
\end{equation}
where $\Lambda=\sqrt{iv^2\mcT/\vep}$.
In the large time asymptotic with $\vep\mcT\gg 1$, the main contribution to quantum amplitude is due to the term
\begin{equation}\label{q20}
f(v,\mcT)\simeq e^{-\frac{v}{\vep}}\sum_{n=0}^{\infty}\frac{1}{n!}\Big(\frac{v}{\vep}\Big)^n
\sum_{m=0}^{n}C_m^n\Big(\frac{v}{4i}\Big)^m
\sum_{k=0}^{m-1}\Big(-\frac{4\vep c_{n,m}}{v^2}\Big)^{k}J_n\big(\Lambda\big)\, .
\end{equation}
Then taking into account that the argument of the Bessel function $\Lambda$ is a complex value, 
the Bessel functions grow exponentially with time at $|\Lambda|\gg 1$ \cite{YEL}:
$J_n(\Lambda)\sim \big(2\pi/\Lambda\big)^{1/2}\cos(\Lambda n+\pi/2)$.
Therefore the quantum amplitudes grow in time according the stretch exponential
function
\begin{equation}\label{q21}
f\sim \exp\big(v\sqrt{\mcT/2\vep}\big)\Phi(v,\mcT)=
\exp\left(\sqrt{\beta\gamma t}\right) \tilde{\Phi}(\gamma,t) \, ,
\end{equation}
where $\Phi(v,\mcT)$ is a slow varying and not increasing function
of time\footnote{This result of the Laplace inverse transform can be also
obtained by the stationary phase approximation for the long time asymptotics.
In this case the stationary point of the exponential in integrand
(\ref{q17})  is $u_0=v/2\sqrt{i\vep\mcT}$, which immediately yields
the result of Eq. (\ref{q21}).}, while $\gamma= A_0 I/\hh$.

\section{Quantum CTRW}\label{sec:CTQW}
\def\theequation{\thesection.\arabic{equation}}
\setcounter{equation}{0}

Summarizing results of the previous sections, we admit that for the construction
of the quantum CTRW,
one follows a coarse-graining procedure by lumping a set of quantum states
making a trapping cell, or trapping potential,
which is in complete analogy with the classical CTRW, constructed in
Sec.~\ref{sec:CTRW}. Mapping the density operator $\hat{\rho}$
on the basis of the coherent states $|\mathbf{a}\rgl$ in Eq. (\ref{q2}), such that
$\clP_Q(t)\equiv\clP_Q(\mathbf{a}^*,\mathbf{a},t)=
\lgl\mathbf{a}|\hat{\rho}(t)|\mathbf{a}\rgl$,
one obtains for the quantum density of the probability the following quantum
Liouville equation
\begin{equation}\label{ctqw_1a}
\partial_t\clP_Q(t)=\hat{K}\clP_Q(t)\, .
\end{equation}
Here the quantum ``Liouville'' operator is determined in Eq. (\ref{q7}) and it
contains the same overlapping integrals as its classical counterpart
in Eqs. (\ref{trueWF}) and (\ref{mastereq}). As the result,
the coarse-grained quantum process of the transitions between the trapping
clusters is also a Markov process. Transitions between quantum states inside
the potential do not contribute to the transport (to the spreading of the
initial wave packet in the $x$ coordinate), while transitions between any
states of the different trapping clusters lead to the spreading of the wave
packet in the $x$ coordinate.
The main contribution to this transport in the chain of the localized
AMs is due to the transition between the nearest neighbor clusters and are
determined by the jump length PDF $p_Q(x)$.
The quantum transitions between coherent states inside the clusters leads to the 
traps with trapping times, distributed by
the waiting time PDF $w_Q(t)$. The transition probabilities in the Pauli-
van-Kampen master equation reduces to the classical analogs of the transition
probabilities. Therefore, following the
classical consideration, presented in Sec.~\ref{sec:CTRW}, the quantum distribution
function is governed by the Montroll-Weiss equation (\ref{MW}) in the Fourier-
Laplace space
\begin{equation}\label{ctqw_2a}
\bar{\tilde{\clP}}_Q(k,u)=\hat{\clF}\hat{\clL}\clP_Q=
\frac{1-\tilde{w}_Q(u)}{u}\cdot\frac{\bar{\clP}_0(k)}{1-\bar{p}_Q(k)\tilde{w}_Q(u)}\, .
\end{equation}
As already admitted above $\tilde{p}_Q(k)$ is determined by the transitions between 
the localized states $\tilde{p}_Q(k)\approx 1-\beta A_0R^2k^2$.

\subsection{Waiting time PDF $w_Q(t)$}

To estimate the waiting time PDF $w_Q(t)$ for the quantum trapping cluster, or
``trapping potential'', we are interesting in the large time asymptotics.
The waiting time PDF, has a quantum nature and is determined by the quantum four
modes decay rate, which is the inverse value of the quantum amplitudes
$|f|^2$ in Eq. (\ref{q21}). The probability to find a quantum particle
inside trapping cluster after time $t$ (or no jump
during time interval $(0,t)$) is
$$W_Q(t)=\int_t^{\infty}w_Q(t')dt'=\lgl|f|^{-2}\rgl\, .$$
Here $\lgl|f|^{-2}\rgl$ is the averaged value of the inverse quantum amplitudes.
We take into account that parameter $\gamma$ changes randomly for different
clusters as a function of the random localized states.
We take these value being exponentially distributed
$ \frac{1}{\gamma_0}e^{-\gamma/\gamma_0}$, which yields\footnote{This integration can be
considered as the Laplace transform. Therefore the Tauberian theorem can be
applied to the slow varying function $\tilde{\Phi}(\gamma,t)$.}
$\lgl|f|^{-2}\rgl\propto 1/(1+\sqrt{t/\bt})$, where $\bt=1/2\beta\gamma_0^2$.
Therefore one obtains\footnote{The same asymptotic behavior one obtains by
calculating first a local waiting time PDF $w(t,\gamma)$ and
then averaging it over $\gamma$.}
\begin{equation}\label{ctqw_3}
w_Q(t)\simeq \frac{1}{(t/\bt)^{3/2}}\,.
\end{equation}
which yields $\tilde{w}_Q(u)\approx 1-(u\bt)^{1/2}$. Again, this waiting time
PDF leads to the FFPE (\ref{ffpe}) with the transport exponent $\alpha=1/2$.
The solution of the FFPE is obtained in the form of the Fox
$H$ function, presented in Appendix B. This reads
\begin{equation}\label{qB9}
\clP_Q(x,t)=\frac{1}{\sqrt{D_{\frac{1}{2}}t^{\frac{1}{2}}}}H_{1,1}^{1,0}
  \left[\frac{x^2}{D_{\frac{1}{2}}t^{\frac{1}{2}}} \left|\begin{array}{l c}
   (\frac{3}{4},\frac{1}{2})  \\
    (0,2) \\
    \end{array}\right.\right]\, ,
\end{equation}
and its asymptotic behavior corresponds to the stretched Gaussian exponential
function $\clP_Q(y)\sim \frac{1}{\sqrt{D_{\alpha}t^{\alpha}}}
y^{-(1-\alpha)/(2-\alpha)}e^{-y^{2/(2-\alpha)}}$,
where $y\equiv\frac{|x|}{\sqrt{D_{\alpha}t^{\alpha}}}\gg 1$.
Note that for $\alpha=1$ it corresponds to the normal Gaussian distribution.
This solution eventually leads to quantum subdiffusion of a wave packet spreading
with the MSD
\begin{equation}\label{ctqw_5}
\lgl x^2(t)\rgl\sim \sqrt{t}\, .
\end{equation}
Contrary to classical subdifusion, there is no numerical confirmation of this result.
However, it corroborates to an experimental observation of the optically induced exciton
transport in molecular crystals, which exhibits the intermediate asymptotic subdiffusion
\cite{nature2014} with the experimental transport exponent of the order of
$\sim 0.57$.

\section{Conclusion}\label{sec:concl}
\def\theequation{\thesection.\arabic{equation}}
\setcounter{equation}{0}

The review is concerned with the nonlinear Schr\"odinger equation in
the presence of disorder. The dynamics of an initially localized wave packet is
described in both classical and quantum cases.
In both cases, we obtained a subdiffusive spreading, which is explained in
the framework of a continuous time random walk (CTRW), and it is shown that subdiffusion
is due to the transitions between those Anderson modes (AM)s, which are
strongly overlapped. This overlapping is a common feature of both classical and
quantum dynamics and leads to the clustering with
an effective trapping of the wave packet inside each cluster by
an effective potential.
Therefore, the classical dynamics of the wave packet corresponds to the
CTRW, where the basic mechanism of subdiffusion is an entrapping
of the wave packet with delay, or waiting times
distributed by the power law $w(t)\sim 1/t^{1+\alpha}$ with
$\alpha=1/3$. The trapping mechanism determines the
transport exponent $\alpha$, which is due to the number of AMs
contributed to the overlapping integrals according to Eq.
(\ref{am4}).
Note, that the PDF $\clP(x,t)$ in Eq. (\ref{trueWF})
is the exact distribution, and it corresponds to the CTRW in the long
time and the large scale asymptotics described by the FFPE (\ref{ffpe}).

This classical CTRW consideration can be extended on the wave packet spreading in
the framework of the generalized nonlinear
Schr\"odinger equation
\be\label{gnlse}    %
 i\prt_t\psi=\hH\psi+\beta|\psi|^{2s}\psi\, ,
\ee %
where $s\ge 1$ and the Hamiltonian $\hH$ has a pure point
spectrum with the localized eigenfunctions:
$\hH\Psi_k=\clE_k\Psi_k$. For example, the Hamiltonian describes
Wannier-Stark localization \cite{EminHart}, and the discrete
counterpart of Eq. (\ref{gnlse}) with $s=1$ corresponds to
delocalization in a nonlinear Stark ladder \cite{kolovsky,flach,iom_MMNP}.
Repeating  probabilistic consideration of the CTRW based on the
overlapping integrals $A({\bf k})$ of $2s+1$ eigenfunctions
$\Psi_k(x)$, one obtains that Eq. (\ref{gnlse}) describes
subdiffusion with the transport exponent
\be\label{trexp}   %
\alpha=\frac{1}{1+2s}\, .     %
\ee  %
For different values of $s$, this expression coincides with the numerical results of
Refs. \cite{Skokos2010,flach}. This result also correspond to the
topological approach considered in \cite{MI2015,MI2014}.

The situation with the quantum NLSE differs essentially due the
nonzero commutation rule (\ref{q-commute}). Therefore, performing
$c$-number projection of the Heisenberg equations of motion
on the basis of the coherent states, one obtains a quantum master
equation with the same structure of the Liouville operator as the classical one.
In this construction of the quantum master equation, as a quantum continuous
time random walk (QCTRW), in some extent, we follow a van-Kampen coarse-graining
procedure (for example, see details of the discussion in \cite{kreuzer})
by lumping a set of quantum states making a trapping cluster,
which is in a complete analogy with the classical CTRW construction \cite{Iomin}, 
presented in Sec.~\ref{sec:CTRW}.
The transition probabilities in Eq. (\ref{qmastereq}) correspond
to their classical analogs of the
transition probabilities, since the quantum ``Liouville'' operator determined in Eqs.
(\ref{q7}) and (\ref{K-L}) contains the same overlapping integrals as in the classical
counterpart with $\hh=0$. Transitions between quantum
states inside the clusters do not contribute to the transport (to the spreading of the
initial wave packet in the $x$ coordinate), while transitions between any
states belonging to different clusters lead to the spreading of the wave
packet by changing the $x$ coordinate.
The main contribution to this transport in the chain of the localized
AMs is due to the transitions between the
nearest neighbor clusters, which are determined by the jump length PDF $p_Q(x)$.

Therefore, the dynamics of the wave packet corresponds to the
CTRW, where the basic mechanism of subdiffusion is the entrapping
of the wave packet with delay, or waiting, times
distributed by the power law $w_Q(t)\sim 1/t^{1+\alpha}$.
The trapping mechanism also determines the
transport exponent $\alpha=1/2$, which is due to
the quantum four-mode decay, described by Eq (\ref{q15}).
The four-mode decays determine the leakage probability from
trapping clusters and correspondingly determine
the waiting time PDF $w_Q(t)$.

Careful mathematica analysis of the system (\ref{q15}) has been performed in
Ref.~\cite{bikt1986} in the form of the semiclassical expansion.
These results were also verified in recent publications
\cite{bt2006,bt2008} devoted to the semiclassical analysis of quantum singularities in the
dynamics of a Bose-Einstein condensate trapped in a one-dimensional toroidal geometry and
the semiclassical analysis of the four-wave decay in a quantum chain of oscillators.
As it is shown in \cite{bikt1986,bt2006,bt2008} by asymptotic expansion over $\vep\ll 1$, a convergence of quantum solutions to the corresponding classical solutions exists only for the logarithmic time scale $\sim\log(1/\vep$, and beyond this time quantum processes cannot be violated. Therefore, the quantum decay processes determine the quantum kinetics,
which is defined in the large time asymptotics of the quantum dynamics
and correspondingly determine the kinetic coefficients of the generalized master equation of the quantum kinetics.

It should be noted that iteration equation (\ref{qmastereq}) is exact in
complete correspondence with its classical counterpart (\ref{mastereq}).
However the fractional integral equation (\ref{ffpe}), which is valid for
both classical and quantum NLSE, is obtained in the
diffusion limit in the framework of the CTRW consideration.
The latter consists of two steps of the Markov independent processes.
These are
the trapping with the power law waiting PDF and instant jumps.
The waiting process is a quantum process,
which is described  by the quantum decay equations (\ref{q15}). The instant jumps are
described classically. In this case, we neglect the quantum terms $\sim {\rm O(\hh)}$,
which are the second derivatives in the Liouville operator $\hat{L}_1$.
As discussed in Ref. \cite{bt2008} the dimensionless Planck constant (as a semicalssical parameter) is small $\hh\sim 1/N$ due to the large
number of particles in mesoscopic Bose-Einstein condensate systems
$N\sim 10^{3}\div 10^{6}$, and violation of the quantum terms
$\sim {\rm O}(1/N)$ is well justified for the long distance jumps
described by $p_Q(x)\rightarrow p(x)$.

It is worth mentioning that contrary to classical NLSE (\ref{am1}),
there is no a numerical confirmation of this quantum subdiffusion.
However, an experimental observation of the exciton
transport in molecular crystals exhibits the intermediate asymptotic subdiffusion
\cite{nature2014} with the experimental transport exponent $\alpha\sim 0.57$.

\section*{Acknowledgment}
This research was supported by the Israel Science Foundation
(ISF-1028).

\appendix

\section{A brief survey on fractional integration}
\def\theequation{A.\arabic{equation}}
\setcounter{equation}{0}

Extended reviews of fractional calculus can be found
\textit{e.g.}, in \cite{oldham,podlubny,SKM}. Fractional
integration of the order of $\alpha$ is defined by the operator
\be\label{Ap1}  %
{}_aI_x^{\alpha}f(x)=
\frac{1}{\Gamma(\alpha)}\int_a^xf(y)(x-y)^{\alpha-1}dy\, , \ee %
where $\alpha>0,~x>a$ and $\Gamma(z)$ is  the Gamma function.
Fractional derivation was developed as a generalization of integer
order derivatives and is defined as the inverse operation to the
fractional integral. Therefore, the fractional derivative is
defined as the inverse operator to ${}_aI_x^{\alpha}$, namely $
{}_aD_x^{\alpha}f(x)={}_aI_x^{-\alpha}f(x)$ and
${}_aI_x^{\alpha}={}_aD_x^{-\alpha}$. Its explicit form is
\be\label{Ap2}   %
{}_aD_x^{\alpha}f(x)=
\frac{1}{\Gamma(-\alpha)}\int_a^xf(y)(x-y)^{-1-\alpha}dy\, . \ee %
For arbitrary $\alpha>0$ this integral diverges, and as a result
of this a regularization procedure is introduced with two
alternative definitions of ${}_aD_x^{\alpha}$. For an integer $n$
defined as $n-1<\alpha<n$, one obtains the Riemann-Liouville
fractional derivative of the form
\begin{equation}\label{Ap3}   %
{}_a^{RL}D_x^{\alpha}f(x)\equiv{}_aD_x^{\alpha}f(x)
=\frac{d^n}{dx^n}{}_aI_x^{n-\alpha}f(x)\, ,
\end{equation}
and fractional derivative in the Caputo form
\begin{equation}\label{Ap4}  %
{}_a^CD_x^{\alpha}f(x)= {}_aI_x^{n-\alpha}\frac{d^n}{dx^n}f(x)\, .
\end{equation}  %
There is no constraint on the lower limit $a$. For example, when
$a=0$, one has
${}_0^{RL}D_x^{\alpha}x^{\beta}=\frac{x^{\beta-\alpha}
\Gamma(\beta+1)}{\Gamma(\beta+1-\alpha)} $. This fractional
derivation with the fixed low limit is also called the left
fractional derivative. However, one can introduce the right
fractional derivative, where the upper limit $a$ is fixed and
$a>x$. For example, the right fractional integral is
\begin{equation}\label{Ap5}
{}_xI_a^{\alpha}f(x)=\frac{1}{\Gamma(\alpha)}
\int_x^{a}(y-x)^{\alpha-1}f(y)dy\, .
\end{equation}  %
Another important property is $D^{\alpha}I^{\beta}=I^{\beta-\alpha}$,
where other indexes are omitted for brevity. Note that the inverse
combination is not valid, in general case,
$I^{\beta}D^{\alpha}\neq I^{\beta-\alpha}$, since it depends on
the lower limits of the integrations \cite{podlubny}. We also use
here a convolution rule for the Laplace transform for $0<\alpha<1$
\begin{equation}\label{mt3}
\clL[{}I_x^{\alpha}f(x)]=s^{-\alpha}\tilde{f}(s)\, .
\end{equation}
Note that for arbitrary $\alpha>1$ the treatment of the Caputo
fractional derivative by the Laplace transform is more convenient
than the Riemann-Liouville one.

It should be admitted that solutions, considered here can be
obtained by the Laplace inversion in the form of the
Mittag-Leffler function \cite{batmen,HMS2011,MH2008}
\be\label{Ap7}  %
\clE_{(\nu,\beta)}(zr^\nu)= \frac{r^{1-\beta}}{2\pi
i}\int_{\clC}\frac{s^{\nu-\beta}e^{sr}}{s^{\nu}-z}ds\, , \ee %
where $\clC$ is a suitable contour of integration, starting and
finishing at $-\infty$ and encompassing a circle $|s|\le
|z|^{\frac{1}{\nu}}$ in the positive direction, and $\nu,\beta>0$.

\section{Solution in the form of the Fox $H$ function}
\def\theequation{B.\arabic{equation}}
\setcounter{equation}{0}
The Fox $H$ function is defined in terms of the
Milln-Barnes integral \cite{HMS2011,MH2008,MSH2010}
\begin{equation}\label{B1}
H_{p,q}^{m,n}(z)=
H_{p,q}^{m,n}
\left[ z \left|\begin{array}{l c}
    (a_1,A_1)\,,\dots\, ,(a_p,A_p) \\
    (b_1,B_1)\,,\dots\, ,(b_q,B_q)  \end{array}\right.\right] =
\frac{1}{2\pi i}\int_{\Omega}\Theta(s)z^{-s}ds\,
\end{equation}
where
\begin{equation}\label{B2}
\Theta(s)=\frac{\Big\{\prod_{j=1}^m\Gamma(b_j+sB_j)\Big\}
\Big\{\prod_{j=1}^n\Gamma(1-a_j-sA_j)\Big\}}
{\Big\{\prod_{j=m+1}^q\Gamma(1-b_j-sB_j)\Big\}
\Big\{\prod_{j=n+1}^p\Gamma(a_j+sA_j)\Big\}}
\end{equation}
with $0\le n\le p$, $1\le m\le q$ and $a_i\, , b_j \in C$, while  $A_i\, ,B_j\in R+$,
for $i = 1\,,\dots\, ,p$, and $j = 1\, , \dots\, , q$. The contour $\Omega$ starting at
$\sigma-i\infty$ and ending at $\sigma+i\infty$, separates the poles of the
functions $\Gamma(b_j +sB_j)$, $j = 1\,,\dots\, , m$ from those of the function
$\Gamma(1-a_i -sA_i)$, $i = 1\, ,\dots\, , n$.

Now the Montroll-Weiss equation (\ref{MWasy}) can be solved in terms of the Fox $H$
functions. Let us present the Montroll-Weiss equation (\ref{MWasy}) in the form
\begin{equation}\label{B3}
\bar{\tilde{\clP}}(k,u)=
\frac{u^{\alpha-1}}{u^{\alpha}+D_{\alpha}k^2}\, ,
\end{equation}
where we take $\bar{\tilde{\clP}}_0(k,u)=1$.
Then employing formula (\ref{Ap7})
for the Mittag-Leffler function \cite{podlubny,batmen,HMS2011,MH2008} one obtains
\begin{equation}\label{B4}
\bar{\clP}(k,t)=\clE_{(\alpha,1)}\big(-D_{\alpha}k^2t^{\alpha}\big)\, .
\end{equation}
The two parameter Mittag-Leffler function (\ref{B2}) is a special case of the
Fox $H$-function \cite{HMS2011,MH2008}, which can be represented by means
of the Milln-Barnes integral (\ref{B1})
\begin{eqnarray}\label{B5}
\clE_{(\alpha,\beta)}(-z) =\frac{1}{2\pi i}\int_{\Omega}
\frac{\Gamma(s)\Gamma(1-s)}{\Gamma(\beta-\alpha s)} z^{-s}ds &=&
H_{1,2}^{1,1}
\left[ z \left|\begin{array}{l}
    (0,1) \\
    (0,1), (1-\beta,\alpha)
  \end{array}\right.\right] \nonumber \\
 &=&
\frac{1}{\delta} H_{1,2}^{1,1}
\left[ z \left|\begin{array}{l}
    (0,1/\delta) \\
    (0,1), (1-\beta,\alpha/\delta)
  \end{array}\right.\right]   \, .
\end{eqnarray}
Fourier-cosine transform of Eqs. (\ref{B4}) and (\ref{B5}) yields  \cite{MH2008}
\begin{eqnarray}\label{B6}
\clP_{\rho}(x,t) &=& \frac{1}{2\pi}\int_0^{\infty}dk k^{\rho-1}\cos(kx)H_{1,2}^{1,1}
\left[ \sqrt{D_{\alpha}t^{\alpha}}|k| \left|\begin{array}{l}
    (0,1/2) \\
    (0,1), (0,\alpha/2)
  \end{array}\right.\right]  \nonumber \\
&=&  \frac{1}{|x|^{\rho}}H_{3,3}^{2,1}
  \left[\frac{x^2}{D_{\alpha}t^{\alpha}} \left|\begin{array}{l c}
  (1,1), (1,\alpha), (\frac{1+\rho}{2},1) \\
    (1,2), (1,1), (\frac{1+\rho}{2},1)\\
    \end{array}\right.\right]
\end{eqnarray}
For $\rho=1$ one obtains the solution of Eq. (\ref{MWasy}).
However, taking into account the properties of the Fox H function
\cite{MSH2010}, one obtains
\begin{equation}\label{B7}
\frac{1}{|x|}H_{3,3}^{2,1}
  \left[\frac{x^2}{D_{\alpha}t^{\alpha}} \left|\begin{array}{l c}
  (1,1), (1,\alpha), (1,1) \\
    (1,2), (1,1), (1,1)\\
    \end{array}\right.\right]=
\frac{1}{|x|}H_{2,2}^{2,0}
  \left[\frac{x^2}{D_{\alpha}t^{\alpha}} \left|\begin{array}{l c}
   (1,\alpha), (1,1)  \\
    (1,2), (1,1) \\
    \end{array}\right.\right]
\end{equation}
Then using property $x^{\delta}H_{p,q}^{m,n}\left[x
\left|\begin{array}{l c}    (a_p,A_p)  \\     (b_q,B_q) \\    \end{array}\right.\right]=
H_{p,q}^{m,n}\left[x \left|\begin{array}{l c}    (a_p+\delta A_p,A_p)  \\
(b_q+\delta B_q,B_q) \\     \end{array}\right.\right]$, Eq. (\ref{B6})
reduces to
\begin{equation}\label{B8}
\clP(x,t)=\frac{1}{\sqrt{D_{\alpha}t^{\alpha}}}
H_{2,2}^{2,0}
  \left[\frac{x^2}{D_{\alpha}t^{\alpha}} \left|\begin{array}{l c}
   (1-\frac{\alpha}{2},\alpha), (\frac{1}{2},1)  \\
    (0,2), (\frac{1}{2},1) \\
    \end{array}\right.\right]
\end{equation}
Again using property of Eq. (\ref{B7}), one obtains
\begin{equation}\label{B9}
\clP(x,t)=\frac{1}{\sqrt{D_{\alpha}t^{\alpha}}}H_{1,1}^{1,0}
  \left[\frac{x^2}{D_{\alpha}t^{\alpha}} \left|\begin{array}{l c}
   (1-\frac{\alpha}{2},\alpha)  \\
    (0,2) \\
    \end{array}\right.\right]\, .
\end{equation}

\section{Inferring Eq. (\ref{q14}) for functions $f_p$ and $\tilde{f}_{-p}$}
\def\theequation{C.\arabic{equation}}
\setcounter{equation}{0}

Let us rewrite Eqs. (\ref{q11}) and (\ref{q12}) in the action-angle variables
$(I_k,\varphi_k)$, where we use the variable change
\begin{equation}\label{C1}
\sqrt{\hh}a_k=\sqrt{I_k}\exp({i\varphi_k})\equiv \sqrt{I}\exp(i\varphi)\, .
\end{equation}
Therefore, functions $f_p$ and $\tilde{f}_{-p}$ can be presented in the form
\begin{eqnarray}\label{C2}
f_p &=& f(I,\varphi,t)=\frac{1}{\sqrt{2\pi}}\sum_{n=-\infty}^{\infty}
f_n(I,t)e^{in\varphi}\, , \nonumber \\
\tilde{f}_{-p} &=& g(I,t)=\frac{1}{\sqrt{2\pi}}\sum_{n=-\infty}^{\infty}
g_n(I,t)e^{in\varphi}\, .
\end{eqnarray}
Using relations of differentiations
\begin{eqnarray}\label{C3}
&a_k\frac{\prt}{\prt a_k}=I\frac{\prt}{\prt I}-\frac{i}{2}\frac{\prt}{\prt\varphi}\, ,
\nonumber \\
& a_k^2\frac{\prt^2}{\prt a_k^2}+2|a_k|^2a_k\frac{\prt}{\prt a_k}=
2i\left(\frac{1}{2}\frac{\prt}{\prt\varphi}-
I\frac{\prt}{\prt I}\frac{\prt}{\prt\varphi}-I\frac{\prt}{\prt\varphi}\right)\, ,
\end{eqnarray}
one obtains from Eqs. (\ref{q11}),(\ref{q12}) and (\ref{C2}) the following system
of equations
\begin{equation}\label{C4a}
i\prt_tf_n=(n\omega_k+\omega_{k+p})f_n+\hh\beta A_0[(2+n)I\prt_If_n
+(n/2+nI/\hh+2I/\hh)f_n-Ig_{n-2}/\hh]\, ,
\end{equation}
\begin{equation}\label{C4b}
i\prt_tg_n=(n\omega_k-\omega_{k-p})g_n+\hh\beta A_0[-I(2-n)\prt_Ig_n
+(n/2-2I/\hh+nI/\hh)g_n+If_{n-2}/\hh]\, .
\end{equation}
Due to the initial conditions
$f_p(0)=1$ and $f_{-p}(0)=0$, the initial conditions for the system of Eqs.
(\ref{C4a}),(\ref{C4b}) are $f_n(0)=\delta_{n,0}$ and $g_n(0)=0$. Therefore
the solutions of Eqs. (\ref{C4a}),(\ref{C4b}) are
\begin{equation}\label{C5}
f_n(t)=0 ~~\mbox{for $n\neq 0$}\,, ~~~~~g_n(t)=0 ~~~\mbox{for $n\neq 2$}\, .
\end{equation}
For the functions $f_0(t)$ and $g_2(t)$, one performs the following substitution
\begin{eqnarray}\label{C6}
f_0(I,t)&=&\exp[-i(2\omega_k-\omega_{k-p}+\hh\beta A_0)t]f(I,t)\, ,  nonumber \\ g_2(I,t)&=&\exp[-i(2\omega_k-\omega_{k-p}+\hh\beta A_0)t]g(I,t)\, ,
\end{eqnarray}
and obtains Eqs. (\ref{C4a}) and (\ref{C4b}) in the form
\begin{eqnarray}\label{C7}
i\prt_tf &=& (\omega_{k+p}+\omega_{k-p}-2\omega_k) f+
\hh\beta A_0I \prt_I f +2\hh\beta A_0I f -\beta A_0 I g\, , \nonumber \\
i\prt_tg &=&  \beta V_0I f\, .
\end{eqnarray}

\section{Inverse Laplace transform of Eq.(\ref{q17})}
\def\theequation{D.\arabic{equation}}
\setcounter{equation}{0}

To perform the Laplace inversion in Eq. (\ref{q17})
let us present $f_u(y)$ as follows
\begin{equation}\label{D1}
f_u(v)=-ie^{\frac{v}{\vep}}\sum_{n=0}^{\infty}\frac{1}{n!}\sum_{m=0}^n
C_m^nG_{m,n}(u)\Big(-\frac{v}{4i}\Big)^m\, .
\end{equation}
Here
\begin{equation}\label{D2}
G_{m,n}(u)=\sum_{k=0}^{\infty}\frac{1}{k!}\Big(\frac{v^2}{4i\vep}\Big)^k
\frac{1}{u-ic_{n,m}}\cdot\frac{1}{u^{m+k}}\, ,
\end{equation}
where $c_{n,m}=2-2\vep(m+n)$. Thus, the Laplace inversion reduces to
the integrals in $G_{m,n}(u)$
\begin{equation}\label{D3}
\chi_{n,m,k}=\frac{1}{2\pi i}\int_{\sigma-i\infty}^{\sigma+i\infty}
\frac{e^{u\mcT}du}{(u-ic_{n,m})u^{m+k}}\, .
\end{equation}
The standard residue rules yield
\begin{equation}\label{D4}
\chi_{n,0,0}=e^{ic_{n,0}\mcT}\, ,~~~\chi_{n,m,k}=\frac{e^{ic_{m,n}\mcT}}
{(ic_{m,n})^{m+k}}-\frac{e^{ic_{m,n}\tau}\Gamma(m+k,ic_{m,n})}{ic_{m,n}\Gamma(m+k)}\, ,
\end{equation}
where $\Gamma(l)=(l-1)!$ is a gamma function, while $\Gamma(l,z)$ is an incomplete
gamma function and $1/\Gamma(0)=0$. Note that $\vep$ is such that $c_{n,m}\neq 0$
for all $n$ and $m$.
Taking into account Eqs. (\ref{D2}),(\ref{D3}),(\ref{D4}), the inverse Laplace
in Eq. (\ref{D1}) yields function $f(v,\mcT)$ as follows
\begin{equation}\label{D5}
f(v,\mcT)=e^{-\frac{v}{\vep}}\sum_{n=0}^{\infty}\frac{1}{n!}\Big(\frac{v}{\vep}\Big)^n
\sum_{m=0}^{n}C_m^n\Big(\frac{v}{4i}\Big)^mF_{m,n}(\mcT)\, ,
\end{equation}
where
\begin{eqnarray}\label{D6}
F_{m,n}(\mcT) &=& \frac{e^{ic_{n,m}\mcT}}{(ic_{n,m})^m}
\sum_{k=0}^{\infty}\frac{1}{k!}\Big(\frac{v^2}{4i\vep}\Big)^k
\frac{\Gamma(m+k)-\Gamma(m+k,ic_{n,m}\mcT)}{(ic_{n,m})^k\Gamma(m+k)}
\nonumber \\
&=& \frac{1}{(ic_{n,m})^m}\Big[e^ce^b-\sum_{k=0}^{\infty}\frac{b^k}{k!}
\sum_{l=0}^{m+k-1}\frac{c^l}{l!}\Big]\, ,
\end{eqnarray}
where $b=v^2/4\vep c_{n,m}$ and $c=ic_{n,m}\mcT$.
Here we used an explicit expression for the gamma function $\Gamma(k,z)=e^{-z}(k-1)!
\sum_{l=0}^{k-1}\frac{z^l}{l!}$ and accounted that $1/\Gamma(0)=0$.
Then we change the index in $k=n+l$ in the sums that yields
\begin{eqnarray}\label{D7}
\sum_{k=0}^{\infty}\frac{b^k}{k!}\sum_{l=0}^{m+k-1}\frac{c^l}{l!}\Big]
&=&\sum_{n=0}^{\infty}\sum_{l=0}\frac{b^{n+l}}{(n+l)!}\cdot\frac{c^l}{l!}
+\sum_{k=0}^{\infty}\sum_{n=0}^{m-1}\frac{b^k}{k!}\cdot\frac{c^{k+n}}{(k+n)!}
\nonumber \\
&=& \sum_{n=1}^{\infty}b^n\sum_{l=0}^{\infty}\frac{(-bc)^l}{(n+l)!l!}
+\sum_{n=0}^{m-1}c^n\sum_{l=0}^{\infty}\frac{(-1)^l(-bc)^l}{(n+l)!l!}
\nonumber \\
&=& \sum_{n=1}^{\infty}c^{-n}J_n\big(2\sqrt{-bc}\big)
+\sum_{n=0}^{m-1}b^{-n}J_n\big(2\sqrt{-bc}\big)\, .
\end{eqnarray}
Here we used the definition of the Bessel function \cite{YEL}
$J_n(2z)=\sum_{l=0}^{\infty}\frac{(-1)^l}{(n+l)!l!}z^{2l}$

\end{document}